# Development of a Evaluation Tool for Age-Appropriate Software in Aging Environments: A Delphi Study

## Abstract

**[Background]:** In recent years, the size of the elderly population has further expanded. At the same time, the development of science and technology also poses new challenges and opportunities to the development of the elderly care industry. Software aging has become a key tool to promote digital inclusion and social participation of older people, and despite policy support, there are still problems exist, such as inadequate design, using unfriendly, and missing standards and assessments.

**[Objective]:** We aimed to develop a dependable reliable tool for assessing software age-appropriateness.

**[Methods]:** We conducted a systematic review to get the indicators of technology age-appropriateness from studies from January 2000 to April 2023. This study engaged 25 experts from the fields of anthropology, sociology, and social technology research across, three rounds of Delphi consultations were conducted. Experts were asked to screen, assess, add and provide feedback on the preliminary indicators identified in the initial indicator pool.

**[Result]:** We found 76 criteria for evaluating quality criteria was extracted. , grouped into 11 distinct domains. After completing three rounds of Delphi consultations, experts drew upon their personal experiences, theoretical frameworks, and industry insights to arrive at a three-dimensional structure for the evaluation tool:user experience, product quality, and social promotion. These metrics were further distilled into a 16-item scale, and a corresponding 21-question questionnaire was formulated. The developed tool exhibited strong internal reliability (Cronbach's Alpha = 0.867) and content validity (S-CVI = 0.93).

**[Conclusion]:** This tool represents a straightforward, objective, and reliable mechanism for evaluating software's appropriateness across age groups. Moreover, it offers valuable insights and practical guidance for designing and developing of high-quality age-appropriate software, and assisst age groups to select software they like.

## Introduction

In recent years, there has been a surge in global smartphone usage, with the number of active mobile devices exceeding 10 billion[1]. Notably, smartphone adoption among the older adults is also experiencing an uptick, and the average user now operates more than 45 distinct applications[2]. Relative to younger demographics, older adults are more susceptible to the digital divide[3, 4], and have a constrained capacity to utilize digital technologies[5, 6]. In addition, with economic growth, people's demand for elderly care services is more biased towards professional and high-tech solutions. Price, user friendliness, self-efficacy, and expected benefits became the main influencing factors, which limit their full embrace, of digital technologies of smart elderly care service[7, 8].

China has introduced policies[9, 10] to promote the digital elderly care industry with a focus on creating smart elderly care as a new sector. These initiatives aim to balance commonality with individuality, modernity with tradition, and science with humanity. The ultimate goal is to leverage technology to enhance elderly care services and help seniors bridge the digital divide[8]. Age-appropriate transform of Apps is also an important part of this, but by 2022, fewer than 0.1‰ of publicly available software in China have undergone age-appropriate modifications[11, 12] . This gap signals an urgent need for mechanisms that can swiftly identify and evaluate high-quality, age-appropriate software. Against this backdrop, the present study aims to develop a rapid, generalized tool for assessing the age-appropriateness of software applications, guided by the framework of "social science and technology" theory.

Current approaches to software evaluation predominantly focus on aspects of usability and quality. These evaluations are often directed at mHealth applications. For instance, Handel assessed 35 health-related software applications using a five-metric framework comprising ease of use, reliability, quality, scope of information, and aesthetics, although the rationale behind choosing these particular metrics was not explicitly detailed[13]. Aiguo Wang et al. proposed a classification model for mobile health apps from three dimensions, combining health status (i.e., physical and/or mental health status), health care processes (which health care procedures to choose, disease prevention or management), and PPM (which factors affect behavior change) to classify software into physical health/mental health, Manage/prevent, induce/empower and other multiple categories to assist program developers to quickly identify the distribution of existing applications[14].Similarly, Stoyanov et al. developed the Mobile App Rating Scale (MARS) to evaluate mental health software. The MARS framework simplifies the evaluation process and focuses on five key criteria: engagement, functionality, aesthetics, quality of information, and subjective quality[15].

Adding another layer to the complexity of app evaluations, Xiaoyu Zhang and colleagues have proposed an evaluation index specifically targeted at age-appropriateness for mobile government applications. This index integrates four dimensions: system usability, inclusiveness, emotional support, and service effectiveness[16]. These dimensions add nuance to the evaluation process, acknowledging that age-appropriateness in software design involves more than just usability; it also includes elements that address the emotional and practical needs of the older adult population.

It is evident from previous examples that conventional software evaluation metrics predominantly focus on traditional factors such as ease of use and aesthetics. However, these frameworks often overlook broader social implications, notably the impact of software innovation on the social systems affecting the older adults. The notion of "age-appropriateness" aligns closely with the principles advocated by "Social Science and Technology," which emphasize a human-centrism approach, interdisciplinary integration, and social inclusivity[17]. Accordingly, the development of age-appropriate assessment tools should be guided by "social technology" theory. This approach will integrate humanistic and ethical considerations into the evaluation framework, explore the role of technological innovation in advancing social equity, and examine policy implications as well as transformations in social systems.

The proposed Social Technology for Aging Environment assessment tool, developed within this conceptual framework, aims to transcend conventional metrics by considering the more profound and extensive impacts of technology on older adult individuals' lives. This tool aims to cater to the diverse needs of the older adults, enhance their quality of life, foster the growth of the aging industry, and amplify social inclusiveness in the digital era. Doing so allows older adults to integrate more seamlessly into the digital society. Concurrently, the development of this assessment system offers valuable insights that can guide future efforts in age-appropriate software transformation.

## Objective

This study aims to develop a reliable multidimensional tool to evaluate social technology in aging environments.

## Methods

### Scale Development for Rating Age-Appropriateness of Software

To develop a robust scale for rating software based on its age-appropriateness, this study conducted an exhaustive literature review focusing on evaluation criteria for quality criteria. The search spanned publications from January 2000 to April 2023 (see Box 1 for search strategy details). Studies meeting the following inclusion criteria were considered:

.The primary focus of the research is on software.

.The scope of the research includes the development and validation of evaluation tools, such as quality criteria indices, rating scales, and assessment guidelines.

.Policy documents that directly relate to quality criteria assessment.

From the gathered literature, the research team extracted various software evaluation criteria. Duplicate and irrelevant criteria were subsequently eliminated and will be subjected to further scrutiny through a Delphi consultation.

Box 1 Search strategy.

| | |
|---|---|
| #1 | TI=( App* OR Mobile OR "Smartphone App*" OR "mobile App*" OR software* OR program*) |
| #2 | TI=(Assess* OR Evaluat* OR Appraise OR Check OR Validat* OR Analysis OR Judge) |
| #3 | TI=(Framework OR Tool* OR Questionnaire OR Survey OR Checklist OR Scale* OR Principle OR Guide OR indicator* OR "indicator system") |
| #4 | TI=(develop* OR construct*) |
| #5 | (#1 AND #2) AND #3 AND #4 |

### Expert screening criteria

Mainly divided into five groups of people, (1) elderly service decision makers; (2) Research and development of old society technology; (3) social science and technology researchers; (4) elderly technology implementer; (5) Others

### Three rounds of Delphi

The first round of consultation: the experts are invited to score the importance, judgment basis and familiarity of each indicator on the letter consultation form, and put forward their own opinions and suggestions on the establishment of the indicator system, so as to screen the alternative indicators.

The second round of consultation: only the importance of each indicator is scored, and the mean value of the importance score of each indicator determined by the first round of correspondence is attached to the second round of consultation form for the reference of experts, and the combined weight of the indicator is determined by the expert scoring method and the combined weight product method.

The third round of consultation: continue to improve each indicator, and determine the final indicator system based on the preliminary opinions.

The expert panel for this study was an interdisciplinary team comprised of anthropologists, sociologists, and researchers specializing in aging technologies. This team engaged in a multi-faceted evaluation of potential indicators for the scale, drawing upon theoretical frameworks, practical experience, peer perspectives, and personal insights. Several statistical metrics were calculated for each indicator, including the mean, full-score frequency, coefficient of variation, expert positivity coefficient, authority coefficient (Cr), and coordination coefficient (Kendall's W). These metrics were used to screen potential indicators based on the following criteria:The mean should be at least "mean - 2* standard deviation." The full-score frequency should be at least "full-score frequency - 2* standard deviation."·The coefficient of variation should not exceed "coefficient of variation + 2* standard deviation."

We included indicators that satisfied all the specified criteria and excluded those that fell short. Additionally, we assessed metrics such as the expert positivity coefficient, authority coefficient, and coordination coefficient to gauge the panel's grasp of the project and the consensus among its members. The final step involved assigning weights to the included indicators using the Analytic Hierarchy Process (AHP), ensuring that the sum of the weights for indicators at each hierarchical level equaled one.

### Validity verification of software the age-appropriateness rating scale

To assess the validity of the developed Software Age-Appropriate Degree Rating Scale, this study employs two key metrics: intrinsic reliability and content validity.

(1)Intrinsic Reliability: This metric evaluates whether the items within a domain coherently measure the same concept and share internal consistency. We assessed the scale's internal reliability by calculating Cronbach's α coefficient with SPSS 25.0. A coefficient value of ≥0.7 is deemed indicative of acceptable scale consistency. Additionally, the corrected item-total correlation was computed to determine the necessity of item deletion. Items with a correlation value less than 0.3 were considered for removal to improve the scale's internal reliability. Subsequent to any deletions, the Cronbach's α coefficient was recalculated to assess the impact on reliability.

(2)Content Validity: This metric quantifies the extent to which each item is representative of the domain it is intended to evaluate. Office Excel 16.60 was used to calculate the Content Validity Index (CVI), both at the item level (I-CVI) and scale level (S-CVI). During the scale's construction, a panel of raters and domain experts assessed the content validity of each item, as well as the scale as a whole, providing recommendations for modifications, deletions, or additions as necessary. Items were rated for their importance on a scale of 1 to 7, where 1 represents minimal importance and 7 represents extreme importance. The I-CVI is calculated as the proportion of raters giving an item a rating of 5 or above, divided by the total number of raters. The S-CVI represents the average I-CVI across all items. A scale is considered to possess good content validity when I-CVI ≥ 0.78 and S-CVI ≥ 0.90; otherwise, modifications are required, and the relevant coefficients recalculated.

## Result

### Construction of Indicator Pool for Software Age-Appropriate Scale

In accordance with the predetermined search strategy, a total of 4,343 articles were identified, of which 3,271 were in English and 1,072 were in Chinese. Following a thorough review and screening process, 27 articles were ultimately selected for inclusion, comprising 18 English articles (67%) and 9 Chinese articles (33%), as delineated in Figure 1. An aggregate of 703 indicators were extracted from these 27 articles. Following data consolidation and categorization, 11 broad classes of indicators emerged. These were further distilled to yield a total of 76 indicators that constitute the initial indicator pool for the study. For a comprehensive overview, refer to Table 1.

### Construction of Age-Appropriate Evaluation Index System for Software

Utilizing the Delphi Method, this study engaged 25 experts from the various fields of anthropology, sociology, and social technology research across three periods in 2023: March 20-28, March 28-April 9, and April 20-May 5. Within these periods, three rounds of Delphi consultations were conducted. Experts were asked to screen, assess, add and provide feedback on the 76 preliminary indicators identified in the initial indicator pool. Based on the collective feedback and questionnaire responses, the indicators underwent iterative revisions and supplements to enhance their relevance and efficacy for evaluating age-appropriateness in software. Detailed information regarding these iterative processes is provided in Table 1.

Table 1 Delphi method expert characteristic list.

|  |  | Frequency | Percent |
|---|---|---|---|
| **Age** |  |  |  |
|  | 31-40 | 4 | 16% |

|  | Frequency | Percent |
|---|---|---|
| 41-50 | 9 | 36% |
| 51-60 | 8 | 32% |
| >60 | 4 | 16% |
| **Identity** | | |
| Geriatric service decision maker | 3 | 12% |
| Research and development of old society technology | 8 | 32% |
| Social technology researcher | 9 | 36% |
| Old technology implementer | 3 | 12% |
| Others | 2 | 8% |
| **Professional title** | | |
| Senior | 9 | 36% |
| Intermediate | 1 | 4% |
| Deputy senior | 8 | 32% |
| Others | 7 | 28% |
| **Working years** | | |
| 5-10 Years | 1 | 4% |
| 11-20 Years | 9 | 36% |
| 21-30 Years | 10 | 40% |
| More than 30 years | 5 | 20% |
| **Educational background** | | |
| PhD | 16 | 64% |
| Master | 5 | 20% |

|  |  | Frequency | Percent |
|---|---|---|---|
| Bachelor | | 4 | 16% |

| No. | Positive coefficient | Ca | Cs | Cr | Kendell'W |
|---|---|---|---|---|---|
| 1 | 100% | 0.8864 | 0.8651 | 0.8757 | 0.135 |
| 2 | 100% | 0.9182 | 0.8498 | 0.8840 | 0.222 |
| 3 | 80% | 0.8769 | 0.8153 | 0.8460 | 0.425 |

### Finalization of Evaluation Dimensions Through Delphi Consultation

After completing three rounds of Delphi consultations, experts drew upon their personal experiences, theoretical frameworks, and industry insights to arrive at a three-dimensional structure for the evaluation tool. This structure encompasses User Experience, Product Quality, and Social Promotion. Within these primary dimensions are eight secondary indicators: Usability, Intelligibility, Cost Consideration, Service Experience, Security, Innovation, Ethics, and Social Integration. The weight coefficient for each indicator was determined based on feedback regarding their relative importance. Detailed weightings are presented in Table 3.

### Specific Indicators and Their Relevance

User Experience: The indicators of Usability, Intelligibility, Cost Consideration, and Service Experience aim to assess the ease with which software can be used, the experience it offers, and the associated costs.

Product Quality: The indicators of Security and Innovation target the quality of the software, considering the software's design optimizations and how they contribute to the user's perception of security and innovative functionality.

Social Promotion: The indicators of Ethics and Social Integration evaluate the societal impact of the software, particularly its role in meeting the unique needs of older users and fostering their social integration.

Additional bonus points can be gained through two supplementary indicators: Compliance and Sociability. These assess the alignment of the software with broader policies and social responsibilities, as well as its adaptability within a multicultural context.

### Evaluation System and Methodology

The evaluation system employs a five-point Likert scale and is designed to cater to two groups: consumers and experts.

Consumer Evaluation: Consumers primarily assess the software across the three dimensions of User Experience, Product Quality, and Social Promotion. Their feedback is coded to generate scores that represent their subjective evaluations of the software's usability, quality, and societal impact.

Expert Evaluation: The experts mainly focus on the additional points, making their assessments based on comprehensive expert opinions and societal impact considerations. Given their expertise and extensive experience, they provide a comprehensive and objective evaluation, thereby enhancing the scientific rigor and accuracy of the results.

By integrating both consumer feedback and expert insights, this assessment method prioritizes user experience while incorporating societal responsibility and professional recommendations. This balanced approach provides a solid basis for continuously refining age-tailored software.

### Software App Evaluation: Reliability and Validity Testing of the STAGE Index System Questionnaire

In our study, we used the STAGE (Software Technology And Geriatric Evaluation) framework to create a questionnaire focused on evaluating software apps designed for older adult users. The questionnaire comprises 21 questions, categorized under 16 index items. A preliminary survey was conducted from July to August 2023, which included 26 older adult software users to test the reliability and validity of the questionnaire.

### Intrinsic Reliability

We employed SPSS to input the scale data and conduct a reliability analysis. The analysis revealed the following:The coefficient values across various index items ranged from 0.455 to 0.742. The highest Cronbach's α coefficient was observed in the "Perceptibility" index, scoring an impressive 0.831, indicating excellent internal consistency.Other indices such as "Usability," "Service Experience," "Security," "Innovation," "Ethics," and "Social Integration" had Cronbach's α values ranging from 0.616 to 0.742, suggesting acceptable levels of consistency.

These findings indicate that the questionnaire demonstrates good reliability, particularly in key areas such as perceptibility and usability. For a detailed breakdown of Cronbach's α values for each index, please refer to Table 2.

By understanding the reliability and validity of the STAGE questionnaire through these initial tests, we aim to refine our approach and ensure that the tool is robust and effectively gauges the user experience and quality criteria targeted at the older adults demographic.

Table 2 The results of intrinsic reliability test.

| Index | Coefficient value | No. | Intrinsic reliability | |
| --- | --- | --- | --- | --- |
| | | | The overall correlation of the corrected terms | The coefficient value after deleting the item |
| **Availability** | 0.742 | 1 | 0.445 | 0.340 |
| | | 2 | 0.406 | 0.363 |
| | | 3 | 0.674 | -.061[a] |
| **Perceptibility** | 0.831 | 4 | 0.787 | 0.708 |
| | | 5 | 0.707 | 0.765 |
| | | 6 | 0.675 | 0.854 |
| Cost consideration | 0.455 | 7 | 0.300 | - |
| | | 8 | 0.300 | - |
| Service experience | 0.820 | 9 | 0.550 | 0.146 |
| | | 10 | 0.467 | 0.263 |
| Security | 0.723 | 11 | 0.558 | 0.154 |
| | | 12 | 0.457 | 0.317 |
| Innovation | 0.719 | 13 | 0.580 | - |
| | | 14 | 0.580 | - |
| Ethics | 0.616 | 15 | 0.532 | 0.383 |
| | | 16 | 0.400 | 0.465 |
| | | 17 | 0.348 | 0.503 |
| Social integration | 0.509 | 18 | 0.486 | 0.233 |
| | | 19 | 0.259 | 0.498 |
| | | 20 | 0.274 | 0.461 |
| | | 21 | 0.222 | 0.500 |
| Total | 0.867 | Coefficient after deleting the corresponding entry | | 0.886 |

*Content Validity Assessment*

This study leveraged an importance scoring system to evaluate the content validity of each index. The dimensions of "Function Learnability" and "Operation Simplicity" emerged as being of paramount importance, both registering an Item-Level Content Validity Index (I-CVI) of 1. This unanimous agreement among experts underscores the critical role these factors play in designing age-appropriate software. They specifically contribute to the ease of use and time-efficiency for older adult users. In addition, "Security" and "Cost" were also highly rated, underscoring a strong alignment of perspectives between experts and users on these key indicators. The importance scores for remaining items consistently exceeded 6 points, and the I-CVI values for all items surpassed the threshold of 0.78. Crucially, the Scale-Level Content Validity Index (S-CVI) for the entire evaluation system stands at an impressive 0.93. Alongside the consistently high I-CVI values for individual indices, this robust S-CVI score attests to the overall strong content validity of the evaluation system.

The high marks across both scale-level and item-level content validity metrics suggest that the evaluation tool is not only methodologically sound but also highly relevant for assessing the appropriateness of software designed for an aging population.

Table 3  Content validity test results.

| Indicators | Items | Content validity | |
|---|---|---|---|
| | | Importance score | I-CVI |
| Availability | Function is easy to learn | 6.85 | 1.00 |
| | Easy to operate | 6.92 | 1.00 |
| Perceptibility | Audio-visual effect | 6.08 | 0.85 |
| | Interactive feedback | 6.46 | 0.92 |
| Cost consideration | Direct cost | 6.46 | 1.00 |
| | Indirect cost | 6.54 | 1.00 |
| Service experience | Needs and values considered | 6.77 | 0.89 |
| | After-sales service | 6.38 | 0.84 |
| Security | Information security | 6.69 | 1.00 |
| | System stability | 6.54 | 1.00 |
| Innovativeness | Functional innovation | 6.08 | 0.85 |
|  | Incentive mechanism | 6.31 | 0.85 |
| **Ethics** | Service | 6.54 | 1.00 |
|  | Special customization | 6.00 | 0.85 |
| **Social influence** | Policy awareness | 6.62 | 0.85 |
|  | Social integration | 6.69 | 1.00 |
| **Total** |  | S-CVI | 0.93 |

## Discussion

### Current Trends in Age-Appropriate Technology

The adaptation of technology for an aging society is a focal point in contemporary research, both domestically and internationally. While countries such as France, Germany, the UK, the US, China, and organizations like the European Commission have rolled out development standards, these predominantly emphasize technical aspects like usability, security, and information quality. They fall short in incorporating the subjective experiences of older adult software users and the broader social impact, revealing a gap in assessing the societal implications of technology transformations.

### Unique Contributions of this Study

This study aims to bridge that gap through its evaluation tool, designed in alignment with the concept of "Technology for an Aging Society." Our approach infuses humanistic and ethical elements into the evaluation framework, focusing on user experience, quality criteria, and social impact. Unlike previous models which are predominantly expert-driven, our evaluation approach places significant emphasis on user feedback, with expert opinions serving as additional input. The composite scores provide a more holistic measure of a software's age-appropriateness.

### Reliability and Validity

The internal reliability test shows that the tool has a high overall Cronbach's α coefficient of 0.867, validating its reliability. Among the sub-dimensions, 'Perceptibility' scored the highest, with a Cronbach's α coefficient of 0.831. This is likely because older adult users rely heavily on sight and hearing when interacting with software, necessitating an objective evaluation of these aspects. The "Service Experience" index also scored highly, indicating its importance in the older adult user's software experience.

## Future Direction

However, the low Cronbach's α scores in cost considerations, ethics, and social integration suggest room for improvement. The sensitivity of the older adults to software costs and potential limitations related to their educational level and expressive abilities may have contributed to these low scores. Adjustments or redesigns may be warranted for specific questions, and any changes must meet two criteria: 1) Retest in a larger sample where the corrected item-total correlations remain under 0.3 or domain Cronbach's α values are less than 0.5; 2) Gain agreement for changes from at least two-thirds of the expert panel. By meeting these criteria, this study aims to continually refine its evaluation tool, aspiring to contribute meaningfully to the burgeoning field of age-appropriate technology.

## Evaluating Validity and Next Steps

The developed assessment tool demonstrates a high content validity of 93%, aligning closely with expert assessments. However, this is an evolving tool subject to further refinement, especially as it is tested on larger sample sizes in real-world applications. Though current data supports the tool's acceptability and validity, challenges may arise when applied more broadly. The research team remains committed to continuous refinement, driven by actual user feedback and evolving needs.

The limitations of this study:The selected indicators are purely based on existing literature, and practical application may yield different results.The small sample size in the preliminary study may introduce errors in the evaluation.The older adult demographic for whom the questionnaire is designed might interpret items differently, potentially skewing results.

Future work by the research team will include On-site inspections to fine-tune the indicator system in real-world settings and further test its reliability and validity.

Evaluations of the age-adaptiveness of various software apps. Development of a comprehensive manual detailing item explanations and operational procedures, supported by relevant training sessions or lectures. Collaboration with domain-specific experts to extend the evaluation framework to hardware products.

The development of this age-appropriate software evaluation tool addresses existing gaps in evaluating software for aging populations. Demonstrating strong reliability and validity, it serves as a useful tool for comprehensive app evaluations, fostering technology adaptation to meet the needs of an aging demographic.

## Acknowledgments

The research was supported by Jiangsu Provincial Science and Technology Innovation Center for Aging Society. We would like to thank Professors Zhang Min, Anning and Chen Hongtu for their assistance in developing the original version of the STAGE Scale.

Thanks to the following experts for their valuable advice on the construction of the evaluation system, the list is as follows (in no particular order)：Wang Jiancheng,ZhangJun,Chen Honglin,Zhang Xiaomin,LiNan,Yu Tingting,Mao Yejuan,Zhang Yingchun,Yu Shuaiyong,Zhang

Yunhong,LiBo,Tao Tianshu,Zhang Kai,Xu Lei.Xue Jinsong,Eiz,Xie Yunhua,Liu Yu,Zhou Lei,Iris Chi,Bo Ruhai,Xu Xiaoyu,Yu Xuan.

## Attachment 1

Questionnaire for Evaluation Index System of Science and Technology for the Older Adults Society.

Hello, dear participant!

We are the working group from the Nanjing University, focusing on creating an Evaluation Index System for an Age-Friendly Society. We're honored to invite you to participate in our survey. The purpose of this questionnaire is to evaluate the age-friendliness of existing software apps, drawing upon your insights and experiences. We're also interested in understanding any challenges you face when using these apps.

What Does "Age-Friendly" Mean?

Being "age-friendly" refers to software products designed and operated while taking into account the unique needs and physiological characteristics of older adults. The goal is not only to make software easily usable for the older adults but also to alleviate fears and barriers that may prevent them from confidently engaging with these technologies. This involves modifying software functionalities, optimizing screen displays, improving the information layout, among other things.

Instructions

The survey consists of 21 questions. Each question will be rated on a scale of 0 to 4, where 0 stands for "Strongly Disagree" and 4 for "Strongly Agree." We encourage you to provide feedback based on your personal experiences and be as objective as possible.

Your valuable input will help us improve the age-friendliness of technology, benefiting society as a whole.

Confidentiality

Please rest assured that all information provided will be kept strictly confidential and will only be used for the purposes of this study.

Let's Begin!

By participating in this survey, you agree to the terms outlined above. Thank you for your time and valuable input!

Basic Information

Your gender:

Male            Female

Your age:

  60-65   65-70   70-75   75-80   More than 80

Which mobile phone software do you usually use? (Multiple choice)

○WeChat  ○Pinduoduo  ○Gaode Map   ○Douyin      ○Toutiao

○Meituan  ○Alipay     ○Himalaya    ○Tencent News  ○Other_____

Your current career status:----------

Do you know about software aging- age-appropriate transformation? [multiple choice]

○Did not know  ○Seldom knew  ○General  ○Known a little  ○Fully understood

Software Aging Level Assessment Scale

## Reference

1. Global smartphone usage report,. 2021.
2. mobile CioC. 2023 silver group digital life insight report. 2023.
3. C UoMLMC. - The Effect of Ageism on the Digital Divide Among Older Adults. 2016:-1. doi: - 10.24966/ggm-8662/100008.
4. van Deursen AJ, Helsper EJ. A nuanced understanding of Internet use and non-use among the elderly. 2015;30(2):171-87. doi: 10.1177/0267323115578059.
5. Hedvicakova M, Svobodova L, editors. Internet Use by Elderly People in the Czech Republic. 2017; Cham: Springer International Publishing.
6. Jimoyiannis A, Gravani M. Digital Literacy in a Lifelong Learning Programme for Adults: Educators' Experiences and Perceptions on Teaching Practices. IJDLDC. 2010 01/01;1:40-60. doi: 10.4018/jdldc.2010101903.
7. Vilpponen H, Leikas J, Saariluoma P, editors. Designing Digital Well-being of Senior Citizens. 2020 13th International Conference on Human System Interaction (HSI); 2020 6-8 June 2020.
8. Chen H, Hagedorn A, An N. The development of smart eldercare in China. The Lancet Regional Health - Western Pacific. 2023 2023/06/01/;35:100547. doi: https://doi.org/10.1016/j.lanwpc.2022.100547.
9. Technology MoIaI. Issued by the Ministry of Industry and Information Technology of the Internet application special action plan to transform aging and barrier-free "notice 2021.
10. Office TG. Issued by The General Office of the State Council General Office of the State Council. On solving the scheme of usino g intelligent technical difficulties in the elderly notice.
11. In 2021 China mobile application market status and usage analysis. 2021.
12. China TcpsgotPsRo. the existing 104 websites and APP completed preliminary optimal aging. 2021.
13. Handel MJ. mHealth (mobile health)-using Apps for health and wellness. Explore (New York, NY). 2011 Jul-Aug;7(4):256-61. PMID: 21724160. doi: 10.1016/j.explore.2011.04.011.

14. Wang A, An N, Lu X, Chen H, Li C, Levkoff S. A classification scheme for analyzing mobile apps used to prevent and manage disease in late life. JMIR mHealth uHealth. 2014 Feb 17;2(1):e6. PMID: 25098687. doi: 10.2196/mhealth.2877.
15. Stoyanov SR, Hides L, Kavanagh DJ, Zelenko O, Tjondronegoro D, Mani M. Mobile App Rating Scale: A New Tool for Assessing the Quality of Health Mobile Apps. 2015;3(1):e27. PMID: 25760773. doi: 10.2196/mhealth.3422.
16. Zhang Xiaoyu, Xue Xiang, ZHU Qinghua, ZHAO Yuxiang. Construction and empirical research of evaluation index system for aging design of mobile government service App. 2023 2023-03-25;44(2):22-30. doi: 10.12154/j.qbzlgz.2023.02.003.
17. Kleinman A, Chen H, Levkoff SE, Forsyth A, Bloom DE, Yip W, et al. Social Technology: An Interdisciplinary Approach to Improving Care for Older Adults. Front Public Health. 2021;9:729149. PMID: 35004562. doi: 10.3389/fpubh.2021.729149.